\documentclass[preprint,12pt]{elsarticle}

\usepackage{amssymb}
\usepackage{amsmath}

\usepackage{graphicx}
\usepackage{braket}
\usepackage{dcolumn}
\usepackage{float}
\usepackage{bm}
\usepackage{xcolor}
\usepackage[normalem]{ulem}
\usepackage{hyperref}

\usepackage[utf8]{inputenc}
\usepackage[english]{babel}
\usepackage[T1]{fontenc}
\usepackage{siunitx}
\hypersetup{
  colorlinks   = true, 
  pdfborder={0 0 0},
  urlcolor     = blue, 
  linkcolor    = black, 
  citecolor   = gray 
}
\usepackage{orcidlink}
\usepackage{caption}
\usepackage{subcaption}

\journal{Helyon}

\begin{document}

\begin{frontmatter}

\title{Dynamics of a quantum polariton vortex: Low excitation scenario}
\affiliation[unalbog]{
organization={Departamento de F\'isica, Universidad Nacional de Colombia},
addresline={111321},
city={Bogot\'a},
country={Colombia}}
\affiliation[udea]{
organization={Instituto de F\'isica, Universidad de Antioquia UdeA},
city={Calle 70 No. 52-21, Medellín},
contry={Colombia}}
\affiliation[jqi]{
organization={Joint Quantum Institute, NIST and University of Maryland},
city={College Park, MD, 20742},
contry={USA}}

\author[unalbog]{J.P. Restrepo Cuartas\,\orcidlink{0000-0002-7430-7393}}

\author[unalbog,udea]{C.A. Fl\'orez-Acosta\,\orcidlink{0009-0008-7825-0778}}

\author[unalbog]{J.D. Rodr\'iguez-Dur\'an}

\author[jqi]{D.G. Su\'arez-Forero\,\orcidlink{0000-0002-2757-6320}}

\author[unalbog]{William J. Herrera\,\orcidlink{0000-0002-8411-9409}}
\author[unalbog]{H. Vinck-Posada\,\orcidlink{0000-0002-4771-1526}}

\date{\today}

\begin{abstract}
Quantum vorticity in polariton systems has been traditionally investigated within the frame of many-body phenomena under the mean-field or coherent approaches. In the present work, we show that the fully quantized picture describes richer dynamics for the vortex core at the quantum coupling limit, where two systems exchange an indivisible excitation. The quantum correlations intrinsic to our formalism account for the emergence of a family of trajectories that differ from the circular paths known in macroscopic vorticity phenomena.  These results indicate that there exists a criterion to differentiate the behavior at the edge between the quantum and classical polariton vortex dynamics.
\end{abstract}







\end{frontmatter}

\section{Introduction}
In the last years, two main research branches in the area of quantum fluids in condensed matter and atomic physics have been formulated: the dynamics of quantum fluids without taking into account their angular momentum~\cite{Pines1966,Leggett2004, Giorgini2008,Carusotto2013} and the quantum dynamics of rotational fluids~\cite{Bewley08,Henn09,Dominici18,Zhao2017}. From the account of their macroscopic coherence, it follows that quantum fluids may only support rotational flow in the form of quantized vortices~\cite{feynman55a,Alperin21}. These vortices,  fundamentally topological, are characterized by a phase rotation of integer steps of $2\pi$ around a phase singularity.  All these exotic phenomena have been theoretically described  and experimentally observed in the context of stationary, harmonically trapped atomic Bose-Einstein condensates~\cite{Shin04,Engels03,Sasaki10}. Additional studies have been carried out in the area of quantum fluids of polaritonic systems~\cite{Dominici2015, Dominici2021a}. On the other hand, phenomena in which all vorticity lies within a single effective core are extensively observed in the laboratory~\cite{fisher96a,fetter01a,Li18}. Among the later studies, the analysis of vorticity in the realm of exciton-polaritonic physics takes on special importance. 

Quantum exciton-polaritons~\cite{Suarez16} are hybrid light-matter quasi-particles. They emerge along the strong coupling regime between a planar semiconductor microcavity mode and a quantum excitation in a quantum well (embedded in the anti-node of a cavity mode)~\cite{deng10a}. In a typical experimental setup, polaritons are optically excited in a semiconductor sample using either continuous or pulsed pumps~\cite{Amo2009,Carusotto2013,Dominici2014}. Moreover, in the strong coupling regime, photons remain trapped in the cavity for a sufficiently long time. They are cyclically absorbed and re-emitted by the active media to create excitons. Once coupled, these multicomponent quasi-particles behave neither as light nor matter. Instead, the two bare components (light and matter) transform into normal, hybrid modes: the upper (UP) and lower polariton (LP)~\cite{Carusotto2013,Amo2009,Marie1999}.

The most common approach to address the vorticity phenomena assumes a macroscopic behavior, i.e., they are considered in the so-called thermodynamic limit where the excitation number goes towards infinity. This framework often relies on Schr\"odinger-like or Gross-Pitaevskii-like equations; two approaches commonly used to describe light-matter interacting quantum systems~\cite{Carusotto2013,deng10a}. These methods agree with the recently reported experimental results~\cite{Dominici2021a}, which do not operate under a low excitation (pumping) regime. 

On the other hand, it has been achieved, recently, the transfer of orbital angular momentum   experimentally; opening the possibility of a controlled excitation of quantum polariton vortices~\cite{Dominici2015,Alperin21, Quinteiro2022,Gnusov2023,Wang2022,Lagoudakis2008,Lagoudakis2009}. Along these lines, more recent challenges in quantized vortices on polaritons encompass exciting phenomena like the continuous, off-resonant injection of angular momentum into an exciton-polariton condensate, using a rapidly rotating optical pump~\cite{ Quinteiro2022,Redondo2022}. Also, the spontaneous vortex cluster formation in a non-resonantly driven polariton condensate~\cite{Sitnik2022}. Additionally, the geometrical-topological structures of the dark states~\cite{Xinghui2022} as well as the imprint of dynamical pseudospin textures associated with Berry curvature on polariton condensates~\cite{Dominici2022}. 

In the present work, we build a framework where we analyze true superpositions of the exciton and cavity mode, i.e., we consider actual linear combinations of the bare states of the components. These superpositions, which usually have definite low excitation numbers---low excitation regime---allow quantum interference phenomena to arise. We show that entanglement, which is the foremost resource for quantum computation and information processing~\cite{Horodecki2009}, is shared between the two components of the polariton quasi-particle~\cite{Ramirez2018}. Therefore, interference occurs because of the formation of entangled states of light and matter. Hence, genuine quantum correlations are present and quantum states associated with vorticity are prepared into the light-matter system.

As it stands, our investigation sharply differs from the standard methods wherein quantum excitations are usually avoided. Moreover,  there is a crucial difference between these two types of studies: in indefinite quantum excitations or high excitation approaches, a mean-field theory can catch up with the physics of the vortex and all the relevant phenomenology of the collective behavior. In contrast, 
in our approach, we explore how a richer family of trajectories for the vortex core arises in the dynamics due to quantum superpositions between light and matter. In particular, we will show that quantum (single excitation) exchange coupling leads to the surge of interference effects on both components' trajectories of the vortex cores. As a result, the vortex core trajectories differ from the typical circular paths, establishing a qualitative criterion to differentiate the crossover behavior between the quantum and the distinct macroscopic well-studied polariton vortex dynamics. Additionally, the present approach is complemented by analyzing some physical magnitudes of each vortex field, like the angular momentum.

The paper is organized as follows: in Sec. II, we introduce the physical system and describe the theoretical model. We discuss essential conceptual differences between our approach and the so-called classical descriptions. In Sec. III, we describe the main results and examine the key features of the vortex trajectories related to entanglement and angular momentum. Likewise, we analyze the structure of the field operator phase concerning the entanglement values. Finally, we summarize the implications of our results  on the dynamics of polariton vortices.

\section{Physical System and Quantum Dynamics}

\subsection{Theory}

\begin{figure}[t]
\centering
\includegraphics[width=0.7\textwidth]{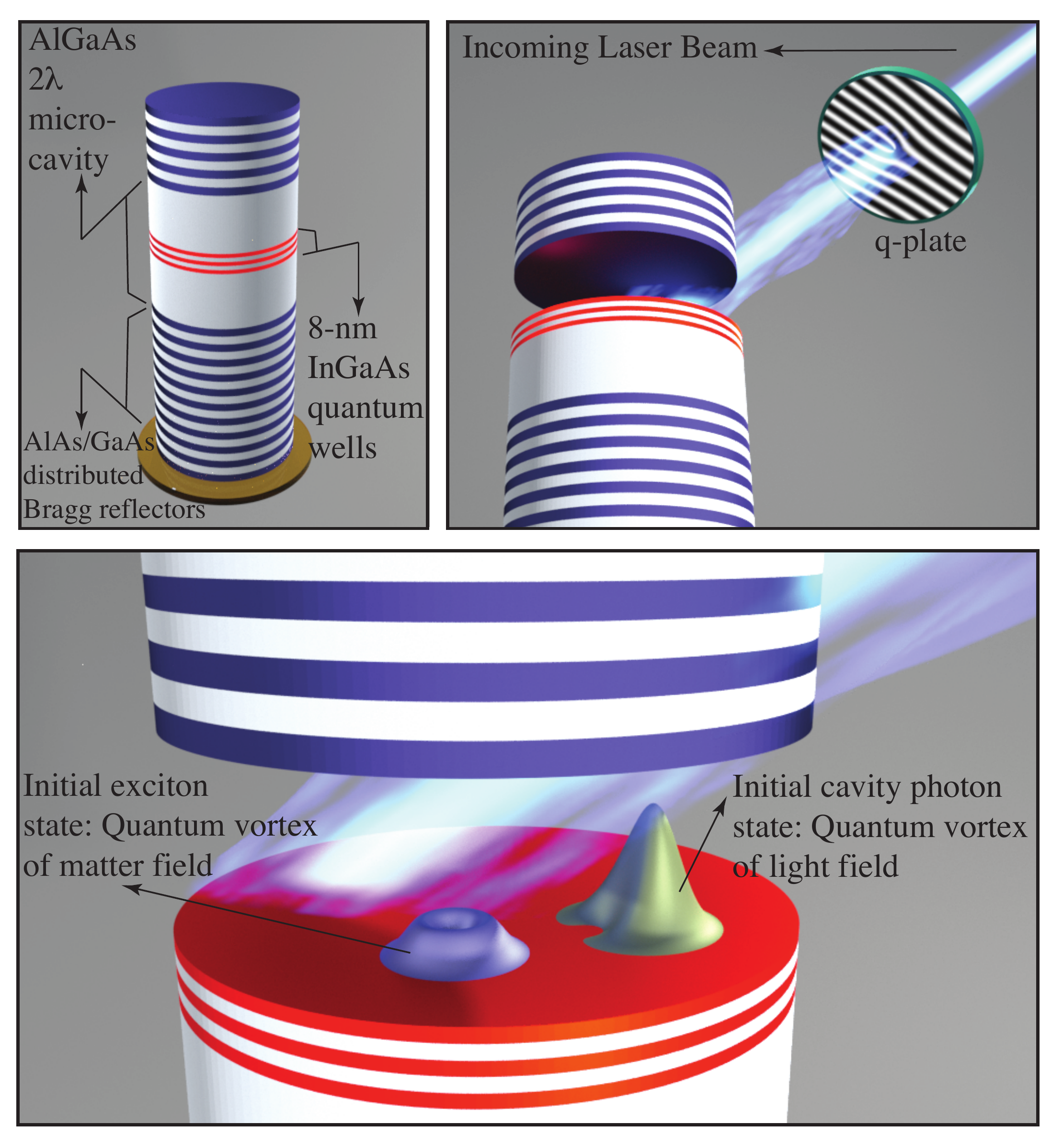}
\caption{Typical polaritonic sample~\cite{Dominici2015}: an AlGaAs microcavity with 
three $8\,\mathrm{nm}$ InGaAs quantum wells at the maximum of the 
cavity field. The cavity is delimited by two distributed Bragg reflectors made of alternated 
AlAs/GaAs layers ($Q = 12,000$). The states are prepared using  an 
80-MHz train of $4.0\,\mathrm{ps}$ laser pulses. The phase shaping 
of the Gaussian pulse is achieved using a q-plate device which performs a wavefront tailoring resulting in a helical vortex structure of the beam~\cite{Petrov2022}.}\label{fig:physys}
\end{figure}

We model a two-dimensional quantum well embedded into a semiconductor microcavity in this contribution. This system has been investigated  experimentally by Dominici et al.~\cite{Dominici18} and is depicted in Fig.~\ref{fig:physys}. In particular, we are interested in the light-matter vortex transference (polariton vortex)~\cite{Fraser2009a,Xuekai2020a}. There are at least  two possible approaches to handle this system: the first is a mean-field approximation reported by Rahmani et al.~\cite{Rahmani2019}; on the other hand, it is possible to make a complete quantum description to obtain the dynamical features of the system. Whereas the first method is valid for specific conditions where the system is composed of a large number of particles close to the thermodynamical limit, the second quantum description captures the quantum correlation properties when the system operates in the few-particles regime. 

We shall start from the  Hamiltonian of the complete   photon-exciton system in the linear regime
\begin{equation}\label{eq:hamiltonian1}
H=
-\frac{\hbar^2}{2\mathrm{m_C}}\nabla^{2} + \mathrm{E_C}  -\frac{\hbar^2}{2\mathrm{m_X}}\nabla^{2} + \mathrm{E_X}+ H_\mathrm{I}\,,
\end{equation}
where  $\nabla^{2}$ is the 2-D Laplace operator and $H_\mathrm{I}$ is the light-matter coupling energy. Here, $\mathrm{m_k}$ and $\mathrm{E_k}$, stand for the effective mass (the gap energy for excitons) and the resonance energy for photons, respectively. Additionally, $\mathrm{k=\{X,C\}}$ stands for the exciton X and the cavity mode C.  Using the common transformation to the second quantization language $p_\mathrm{j}=-(i\hbar/\sqrt{2\mathrm{w}^2})
(O_\mathrm{j}-O_\mathrm{j}^\dagger)$, where $O_{\mathrm{j}}=\{a_{\mathrm{j}}, b_{\mathrm{j}}\}$, $\mathrm{j}=x,y$; and $\mathrm{w}={(\hbar/\mathrm{\omega m_k})}^{\frac{1}{2}}$ is the oscillator length scale. Here we are using the standard nomenclature $a_{\mathrm{j}}$ for photons and $b_{\mathrm{j}}$ for excitons.  Consequently, the Hamiltonian can be written as

\begin{align}\label{eq:hamiltonian2}
H =& 
-\frac{\hbar^2}{4\mathrm{m}_\mathrm{C} \mathrm{w}^{2}}(-2a_x^\dagger a_x+a_x^\dagger a_x^\dagger+a_x a_x-2a_y^\dagger a_y+a_y^\dagger a_y^\dagger+a_ya_y-2)
\\ \nonumber
&- \frac{\hbar^2}{4\mathrm{m}_\mathrm{X} \mathrm{w}^{2}}(-2b_x^\dagger b_x+b_x^\dagger b_x^\dagger+b_x b_x-2b_y^\dagger b_y+b_y^\dagger b_y^\dagger+b_yb_y-2) \\ \nonumber
&+\hbar\Omega (a_x^\dagger b_x+a_x b_x^\dagger+a_y^\dagger b_y+a_y b_y^\dagger) + \mathrm{E_C} +\mathrm{E_X}\, ,
\end{align}

where, by symmetry considerations, our model neglects cross-terms couplings and incorporates uniquely those sharing the same polarization.  

Typical quantum well-microcavity systems are often considered within the wide parabolic approximation. In this approach,  the potential energy for the exciton quasi-particle is described as the one of a harmonic oscillator.~\cite{Vinck-Posada2007, Tabata2007,Tabata2010,Gusev2002}. Furthermore, the photonic field inside an optical microcavity is confined within a small mode volume~\cite{Kasprzak2010}. Therefore, the Hamiltonian can be written in the form 

\begin{align}\label{eq:hamiltonian3}
H=& \ \mathcal{E}_\mathrm{C}(a_x^\dagger a_x+a_y^\dagger a_y)+  \mathcal{E}_\mathrm{X}(b_x^\dagger b_x+b_y^\dagger b_y) 
\\ \nonumber
&+\hbar\Omega (a_x^\dagger b_x+a_x b_x^\dagger+a_y^\dagger b_y+a_y b_y^\dagger)\,,
\end{align}
where $\mathcal{E}_\mathrm{k}=-\frac{\hbar^2}{4\mathrm{m}_\mathrm{k} \mathrm{w^2}} + \mathrm{E}_\mathrm{k}$.

In this work, we explore new features that can be held in the system when the light-matter coupling term exclusively dominates the dynamics. We neglect any non-linear term because we are working in the low excitation scenario, i.e., the low density of excitons does not allow the emergence of non-linearities for stabilization or other effects. Thus, any contribution coming from a term proportional to $\bra{\mathbf{r}}\rho\ket{\mathbf{r}}=|\psi|^2$ in the Hamiltonian will not be considered.

\subsection{Methodology}

The Hamiltonian~(\ref{eq:hamiltonian3}) leads to the equation of motion of the given system. Hence, the time evolution of the polaritonic fluid is ruled by the Liouville–von Neumann (LvN) equation
\begin{equation}\label{eq:liouvillevn}
 \frac{d\rho}{dt} =-\frac{i}{\hbar}[H,\rho]
\end{equation}
where $\rho$ is the density operator for the complete system. A semi-analytical approach to the solution of a pair of coupled harmonic oscillators can be found in~\cite{Kavokin2017}. Our model considers two pairs of quantum harmonic oscillators coupled only in the same direction; for this reason, we use a span method to reckon the dynamical features of the system. 

We start by solving the equation of motion~(\ref{eq:liouvillevn}) by
projecting it to the two-dimensional harmonic oscillator basis $\ket{\mathrm{n_a,m_a}}\otimes\ket{\mathrm{n_b,m_b}}$, where $n_\mathrm{j} (m_\mathrm{j})$ reffers to the quantum numbers associated with the  $x (y)$  polarization, respectively. Then, by using  the Hamiltionian~(\ref{eq:hamiltonian3}) and replacing it in the LvN equation, once we take the matrix elements, we obtain the  set of coupled ordinary differential equations

\begin{align}\label{eq:liouvillevn2}
\dot{\rho}^{\mathrm{n_a,m_a,n_b,m_b}}_{\mathrm{n^\prime_a,m^\prime_a,n^\prime_b,m^\prime_b}}(t)
=& \ 
i\left[\mathcal{E}_\mathrm{C}(\mathrm{n^\prime_a+m^\prime_a-n_a-m_a})\right.\\ \nonumber
&+\left.\mathcal{E}_\mathrm{X}(\mathrm{n^\prime_b+m^\prime_b-n_b-m_b})\right]\rho^{\mathrm{n_a,m_a,n_b,m_b}}_{\mathrm{n^\prime_a,m^\prime_a,n^\prime_b,m^\prime_b}}(t)\\ \nonumber
&+i\Omega\sqrt{\mathrm{n^\prime_b(n^\prime_a+1)}}\rho^{\mathrm{n_a,m_a,n_b,m_b}}_{\mathrm{n^\prime_a+1,m^\prime_a,n^\prime_b-1,m^\prime_b}}(t)\\ \nonumber
&+i\Omega\sqrt{\mathrm{n^\prime_a(n^\prime_b+1)}}\rho^{\mathrm{n_a,m_a,n_b,m_b}}_{\mathrm{n^\prime_a-1,m^\prime_a,n^\prime_b+1,m^\prime_b}}(t)\\ \nonumber
&+i\Omega\sqrt{\mathrm{m^\prime_b(m^\prime_a+1)}}\rho^{\mathrm{n_a,m_a,n_b,m_b}}_{\mathrm{n^\prime_a,m^\prime_a+1,n^\prime_b,m^\prime_b-1}}(t)\\ \nonumber
&+i\Omega\sqrt{\mathrm{m^\prime_a(m^\prime_b+1)}}\rho^{\mathrm{n_a,m_a,n_b,m_b}}_{\mathrm{n^\prime_a,m^\prime_a-1,n^\prime_b,m^\prime_b+1}}(t)\\ \nonumber
&-i\Omega\sqrt{\mathrm{n_b(n_a+1)}}\rho^{\mathrm{n_a+1,m_a,n_b-1,m_b}}_{\mathrm{n^\prime_a,m^\prime_a,n^\prime_b,m^\prime_b}}(t)\\ \nonumber
&-i\Omega\sqrt{\mathrm{n_a(n_b+1)}}\rho^{\mathrm{n_a-1,m_a,n_b+1,m_b}}_{\mathrm{n^\prime_a,m^\prime_a,n^\prime_b,m^\prime_b}}(t)\\ \nonumber
&-i\Omega\sqrt{\mathrm{m_b(m_a+1)}}\rho^{\mathrm{n_a,m_a+1,n_b,m_b-1}}_{\mathrm{n^\prime_a,m^\prime_a,n^\prime_b,m^\prime_b}}(t)\\ \nonumber
&-i\Omega\sqrt{\mathrm{m_a(m_b+1)}}\rho^{\mathrm{n_a,m_a-1,n_b,m_b+1}}_{\mathrm{n^\prime_a,m^\prime_a,n^\prime_b,m^\prime_b}}(t)\,.
\end{align}

According to~\cite{Rahmani2019} (avoiding any external and non-unitary theoretical description of the pumping excitations),  vortices may be introduced  through  initial conditions:
\begin{equation}\label{eq:inwf}
   \braket{\mathbf{r}|\psi_\mathrm{k}(0)}= \frac{e^{-\frac{x^2+y^2}{2\mathrm{w}^2}}Z_\mathrm{k}(x,y,0)}{\mathrm{w}
   \sqrt{\pi } \sqrt{\mathrm{w}^2+|Z_\mathrm{k}(0,0,0)|^2}}\,,
\end{equation}

where $Z_\mathrm{k}(x,y,t)=x-x_\mathrm{k}(t)+i(y-y_\mathrm{k}(t))$. Here, the ordered pair $(x_\mathrm{k}(t),y_\mathrm{k}(t))$ is the time-dependent position of each vortex core in the plane. This  vortex wave function is characterized by the Gaussian spot of size $\mathrm{w}$ which corresponds to the aforementioned oscillator length.

Additionally, we consider the case of a different initial condition which is more related to the coherent nature of the pumping laser. The holographic q-plate imprints the orbital angular momentum to a laser beam better described in the cavity by a coherent state. However, equation~(\ref{eq:inwf}) seems to describe a topological vortex imprinted to the vacuum. Therefore, we draw on an initial condition where a rotational core is applied to a coherent state (see~\ref{app:cohIC}).

 We assume that the two subsystems are not coupled before the vortex state's injection into the system. Then, at $t=0$ the state is entirely uncorrelated, i.e., there is no entanglement between the bipartite system composed of excitons and photons. Therefore, hereafter for all calculations,  we can write the most general disentangled density operator as a convex sum $\rho=\sum_{i}\mathrm{a}_i\ \rho_\mathrm{C}^i(0)\otimes\rho_\mathrm{X}^i(0)$ where $\sum_{i}\mathrm{a}_i=1$, $\mathrm{a}_i\in\mathbb{R}$. We choose the case $\mathrm{a}_1=1$ for simplicity. The initial state of each component is considered as a pure state $\rho=\rho_\mathrm{C}(0)\otimes\rho_\mathrm{X}(0)= \ket{\psi_\mathrm{C}(0)}\bra{\psi_\mathrm{C}(0)}\otimes\ket{\psi_\mathrm{X}(0)}\bra{\psi_\mathrm{X}(0)}$. We can obtain ---from $\rho$, the total density operator--- the reduced density operators for cavity and exciton $\rho_\mathrm{C}=\mathrm{Tr}_\mathrm{X}[\rho]$ and $\rho_\mathrm{X}=\mathrm{Tr}_\mathrm{C}[\rho]$, respectively.    

The state operator can  be spanned for both, exciton and photon, in the coordinate representation by means of the two-dimensional, coordinate projected, basis  $\braket{x,y|n,m}=\braket{\mathbf{r}|n,m}=\mathrm{u_{n,m}}(x,y)$. Here the basis functions are defined as $ \mathrm{u_{n,m}}(x,y)=\mathrm{c_{n,m}H_{n}}(\frac{x}{\mathrm{w}})\mathrm{H_{m}}(\frac{y}{\mathrm{w}})\exp[-(x^2+y^2)/2\mathrm{w}^2]$, where $\mathrm{H_{n}}(z)$ is the Hermite polynomial of degree $n$,  and the normalization constants are  $\mathrm{c_{n,m}=[\pi w^{2}2^{n+m}\Gamma(n+1)\Gamma(m+1)]^{-\frac{1}{2}}}$. The matrix elements of the reduced density operator for the initial condition can be written as 

\begin{subequations}\label{eq:incond}
\begin{align}
        \rho^\mathrm{k}_{0,0,0,0}(0) &=\frac{x^2_\mathrm{k}+ y^2_\mathrm{k}}{\mathrm{w}^2+x^2_\mathrm{k}+ y^2_\mathrm{k}}\,,\\
        \rho^\mathrm{k}_{0,1,0,1}(0)&=\rho^\mathrm{k}_{1,0,1,0}(0) =\frac{1}{2}\frac{\mathrm{w}^2}{\mathrm{w}^2+x^2_\mathrm{k}+ y^2_\mathrm{k}}\,,\\
        \rho^\mathrm{k}_{0,0,0,1}(0) &=\frac{1}{\sqrt{2}}\frac{\mathrm{w}(ix_\mathrm{k}- y_\mathrm{k})}{\mathrm{w}^2+x^2_\mathrm{k}+ y^2_\mathrm{k}}\,,\\
        \rho^\mathrm{k}_{0,0,1,0}(0) &=\frac{1}{\sqrt{2}}\frac{\mathrm{w}(iy_\mathrm{k}- x_\mathrm{k})}{\mathrm{w}^2+x^2_\mathrm{k}+ y^2_\mathrm{k}}\,,\\
        \rho^\mathrm{k}_{0,1,1,0}(0) &=\frac{1}{2}\frac{i\mathrm{w}^2}{\mathrm{w}^2+x^2_\mathrm{k}+ y^2_\mathrm{k}}\,,
  \end{align}
  \end{subequations}
the other matrix elements can be obtained from the condition $\rho^\dagger=\rho$. Elements with any other quantum number are zero. Consequently,  the density matrix of the whole system (polariton) at $t=0$ can be written as  

\begin{equation}\label{eq:globalinitialcon}
\rho^{\mathrm{n_a,m_a,n_b,m_b}}_{\mathrm{n^\prime_a,m^\prime_a,n^\prime_b,m^\prime_b}}(0)=\mathrm{\rho^C_{n_a,m_a,n_a^\prime,m_a^\prime}(0)\rho^X_{n_b,m_b,n_b^\prime,m_b^\prime}(0)}\,.
\end{equation}
With the solution of the equation (\ref{eq:liouvillevn2}) projected into the two-dimensional harmonic oscillator basis $\mathrm{\ket{n_a,m_a}\otimes\ket{n_b,m_b}}$, and using the total initial condition~(\ref{eq:globalinitialcon}), the diagonal density can be spanned by the coordinate representation basis, i.e., the Hermite polynomials in two dimensions

\begin{subequations}\label{eq:span}
\begin{align}
 \mathrm{D_C}(\mathbf{r},t)&=\sum_{\mathrm{\substack{n,m\\n^\prime,m^\prime}}}\mathrm{\rho^C_{n,m,n^\prime,m^\prime}}(t)\mathrm{u_{n,m}}(x,y)\mathrm{u_{n^\prime,m^\prime}}(x,y)\,,\\
\mathrm{D_X}(\mathbf{r},t)&=\sum_{\mathrm{\substack{n,m\\n^\prime,m^\prime}}}\mathrm{\rho^X_{n,m,n^\prime,m^\prime}}(t)\mathrm{u_{n,m}}(x,y)\mathrm{u_{n^\prime,m^\prime}}(x,y)\,.
 \end{align}
 \end{subequations}

The initial condition and parameters entail the core evolution to be restricted among the low excitation manifolds. Hence, the density can be approximated to the general form $\mathrm{D_k}(\mathbf{r},t)=\left(\mathrm{a+b} x+\mathrm{c} y+\mathrm{d} x^2+\mathrm{e} y^2+\mathrm{g} x y\right)\\
\exp[-(x^2+y^2)/\mathrm{2 w^2}] $. Then, we can find the core position  by seeking for the extremum of  the quadratic polynomial $\nabla\left(\mathrm{a+b} x+\mathrm{c} y+\mathrm{d} x^2+\mathrm{e} y^2+\mathrm{g} x y\right)=0$ which yields

\begin{subequations}
\begin{align}
x &\to -\mathrm{\frac{2 b e-c g}{4 d e-g^2}}\,,\\
y &\to -\mathrm{\frac{b g-2 c d}{g^2-4 d e}}\,,
\end{align}
\end{subequations}
where the parameters $\mathrm{a,\, b,\, c,\, d,\, e,\, g\,}$ correspond to combinations of matrix elements 
$\mathrm{\rho^k(n,m,n^\prime,m^\prime)}(t)$.

The main analysis of the results regards the quantum correlations that arise in the system. Accordingly, we use the \textit{linear entropy} to measure the entanglement shared by the excitons and the cavity photons. This measure is defined as 

\begin{equation}\label{eq:linearentropy}
    \mathrm{S_L}(\rho_\mathrm{k})=1-\mathrm{Tr}[\rho_\mathrm{k}^2]\,,
\end{equation}
with $\mathrm{k=C,X}$. The linear entropy measures entanglement because the entire state remains pure in the Hamiltonian dynamics, ruled by the LvN equation~(\ref{eq:liouvillevn}). Similarly, we calculate the angular momentum as a function of time to compare it with the quantum correlations present in the system. The mean value of the angular momentum of the vector state $\psi$ is given by
\begin{equation}\label{eq:angularmomentumfunct}
\braket{L_z}=-i\hbar \int r \mathrm{d}r \mathrm{d}\phi \Braket{\psi|\mathbf{r}}\frac{\partial}{\partial\phi}\Braket{\mathbf{r}|\psi}\,,
\end{equation}
with $r=\sqrt{x^2+y^2}$ and $\phi=\arg(y/x)$ in polar coordinates. 

Nevertheless, in our abstract operator language, the $z$ projection of the angular momentum operator takes the form 

\begin{subequations}\label{eq:amoperator}
\begin{align}
    L^\mathrm{C}_z&=i\hbar(a_x a_y^\dagger-a_x^\dagger a_y)\,,\\
    L^\mathrm{X}_z&=i\hbar(b_x b_y^\dagger-b_x^\dagger b_y)\,.
\end{align}
\end{subequations}

The mean value of these quantities is calculated as usual $\braket{L^\mathrm{k}_z} = \mathrm{Tr}[\rho L^\mathrm{k}_z]$.

\section{Results and Discussion}

\subsection{Quantum Vortex trajectories as a criterion of non-classicality}

\begin{figure}[p]
\centering
\includegraphics[width=0.75\textwidth]{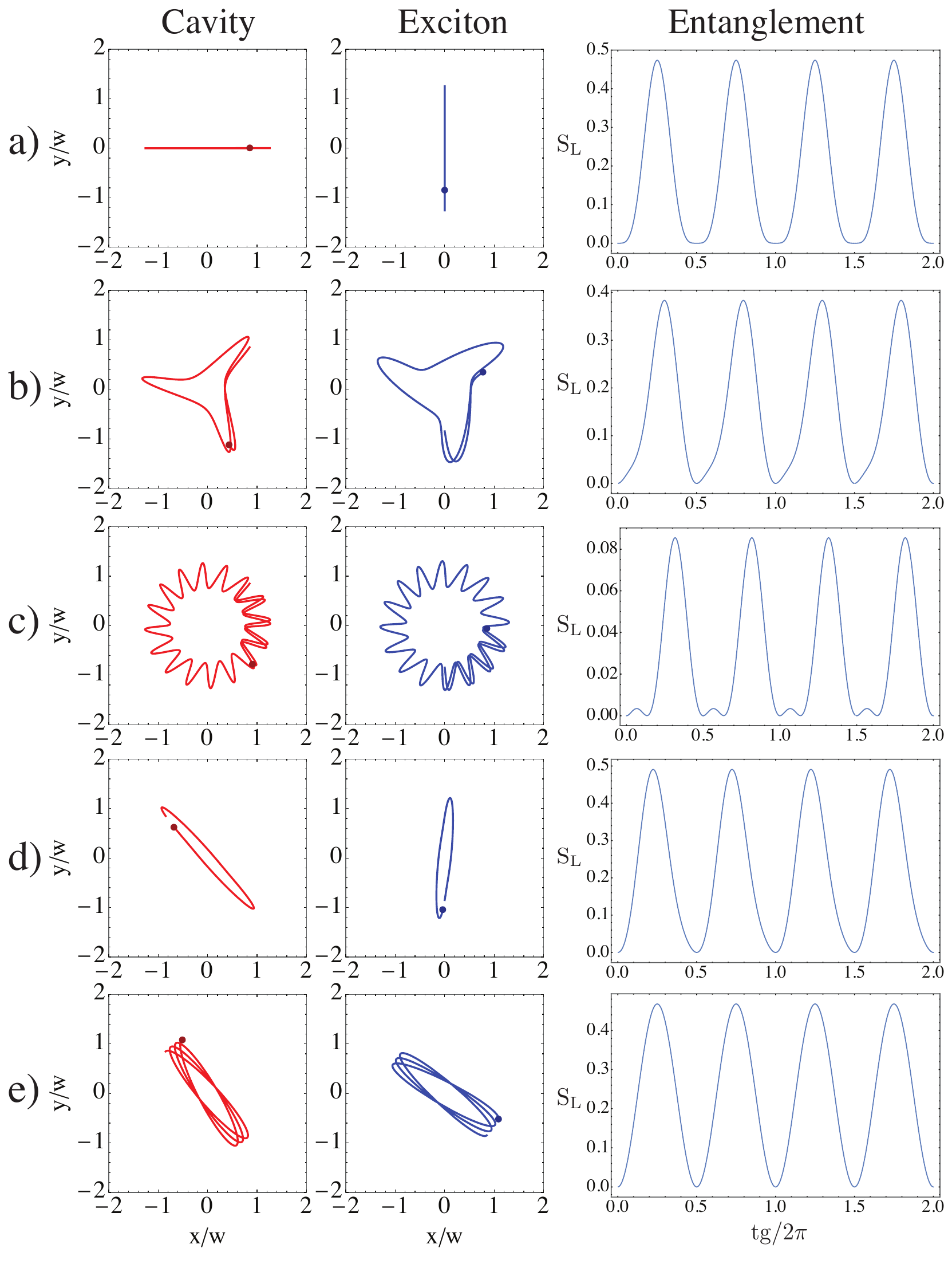}
\caption[Core trajectories and entanglement for different quantum initial conditions and parameters.]{Core trajectories and entanglement for different quantum initial conditions and parameters: $\hbar\Omega = 2.75\, \text{meV}$, $ \mathrm{w} = 25\, \mu\text{m}$, and $\mathrm{L}_{0}=3\sqrt{2}\mathrm{w}/5$.   (a) $\hbar\Delta= 0\, \text{meV}$, $\mathrm{x}_{\mathrm{C}} =\mathrm{L}_{0}$, $\mathrm{y}_{\mathrm{C}} = 0\, \mu\text{m}$, $\mathrm{x}_\mathrm{X} = 0\, \mu\text{m}$, and $\mathrm{y}_{\mathrm{X}} = -\mathrm{L}_{0}$. (b) $\hbar\Delta= 2\, \text{meV}$, $\mathrm{x}_{\mathrm{C}} = \mathrm{L}_{0}$, $\mathrm{y}_{\mathrm{C}} = \mathrm{L}_{0}$, $\mathrm{x}_{\mathrm{X}} = 0\, \mu\text{m}$, and $\mathrm{y}_\mathrm{X} = -\mathrm{L}_{0}$. (c) $\hbar\Delta= 10\, \text{meV}$, $\mathrm{x}_{\mathrm{C}} = \mathrm{L}_{0}$, $\mathrm{y}_{\mathrm{C}} = \mathrm{L}_{0}$, $\mathrm{x}_{\mathrm{X}} = 0\, \mu\text{m}$, and $\mathrm{y}_{\mathrm{X}} = -\mathrm{L}_{0}$. (d) $\hbar\Delta= 0\, \text{meV}$, $\mathrm{x}_{\mathrm{C}} = -\mathrm{L}_{0}$, $\mathrm{y}_{\mathrm{C}} = \mathrm{L}_{0}$, $\mathrm{x}_{\mathrm{X}} = 0\, \mu\text{m}$, and $\mathrm{y}_{\mathrm{X}} = -\mathrm{L}_{0}$. (e) $\hbar\Delta= -0.1\, \text{meV}$, $\mathrm{x}_{\mathrm{C}} = -\mathrm{L}_{0}$, $\mathrm{y}_{\mathrm{C}} = \mathrm{L}_{0}$, $\mathrm{x}_{\mathrm{X}} = \mathrm{L}_{0}$, and $\mathrm{y}_{\mathrm{X}} = -\mathrm{L}_{0}$.}
\label{fig:trajectories}
\end{figure}

This section shows how our full quantum model gives rise to non-trivial trajectories for the vortex core with no classical counterpart. The whole state of the system remains pure, however, the presence of quantum entanglement forbids assigning a vector state to each subsystem. Thus, the reduced state of the photon and exciton is generally a mixed state (this is, in fact, a measure of entanglement via Schmidt rank~\cite{Horodecki2009}) for most of the dynamics. Concerning this,  Fig.~\ref{fig:trajectories} (left and center panel) presents a set of trajectories for the vortex core when different detuning values, $\Delta =(\mathrm{E}_\mathrm{C}-\mathrm{E}_\mathrm{X})/\hbar$, and different initial conditions are considered.

These panels, ranging from a) to e), illustrate how the vortex cores for the cavity photon and the exciton evolve dynamically, following paths that clearly differentiate from those obtained by mean-field approaches. There is always a remnant detuning for $\Delta=0$ because of the different values of the effective masses of the photon and the exciton, i.e,  $\delta=\hbar/ \mathrm{4\mu w^2}$ with $\mu=\mathrm{m_Cm_X/(m_C+m_X)}$ the reduced mass of the components. For this reason, the orbits still precess when the detuning vanishes. 

Along with the core trajectories, we show (right panel) how the linear entropy $\mathrm{S}_{\mathrm{L}}= 1-\text{Tr}(\rho^2)$  behaves as a function of the adimensional time-scale $\mathrm{tg}/2\pi$, where $\mathrm{g}=\sqrt{\Omega^{2}+\Delta^2/4}$ is the dressed (natural) frequency of the compound system. As is seen in these graphs, the entanglement of the component states displays an expected oscillatory behavior as befits to a unitary evolution of a Rabi-like model. Nonetheless, it is quite interesting to observe that in non-trivial trajectories, as observed in panels (a)-(b) and (d)-(e), the entanglement goes from zero (pure state) to its highest value just below the expected numerical value for a maximally mixed reduced density operator; pointing to true quantum dynamics in a time-scale of $tg=2\pi$. However, this behavior becomes less apparent for both components in panel c), where is easily seen how the core moves over a trajectory bounded by two enveloping circles. The entanglement goes progressively to zero as long as the core moves towards the turning points, as shown through the measure of $\mathrm{S}_{\mathrm{L}}$. By contrast, when the core moves away from the turning points, the entanglement slightly increases compared to trajectories with a lower detuning value.
Consequently, the small deviations on this trajectory from the circle can be seen as small quantum corrections to the classical limit. We see this as a consistent result that resembles the circled trajectories described in ~\cite{Rahmani2019} and observed in ~\cite{Dominici2021a}. 

The behavior of quantum entanglement reveals distinct characteristics for different parameters. For example, panels (a) and (c) depict a significant period during which the quantum correlations diminish to zero. Moreover, the oscillation pattern of the linear entropy exhibits asymmetries for specific parameters.

\begin{figure}[t]
\centering
\includegraphics[width=0.66\textwidth]{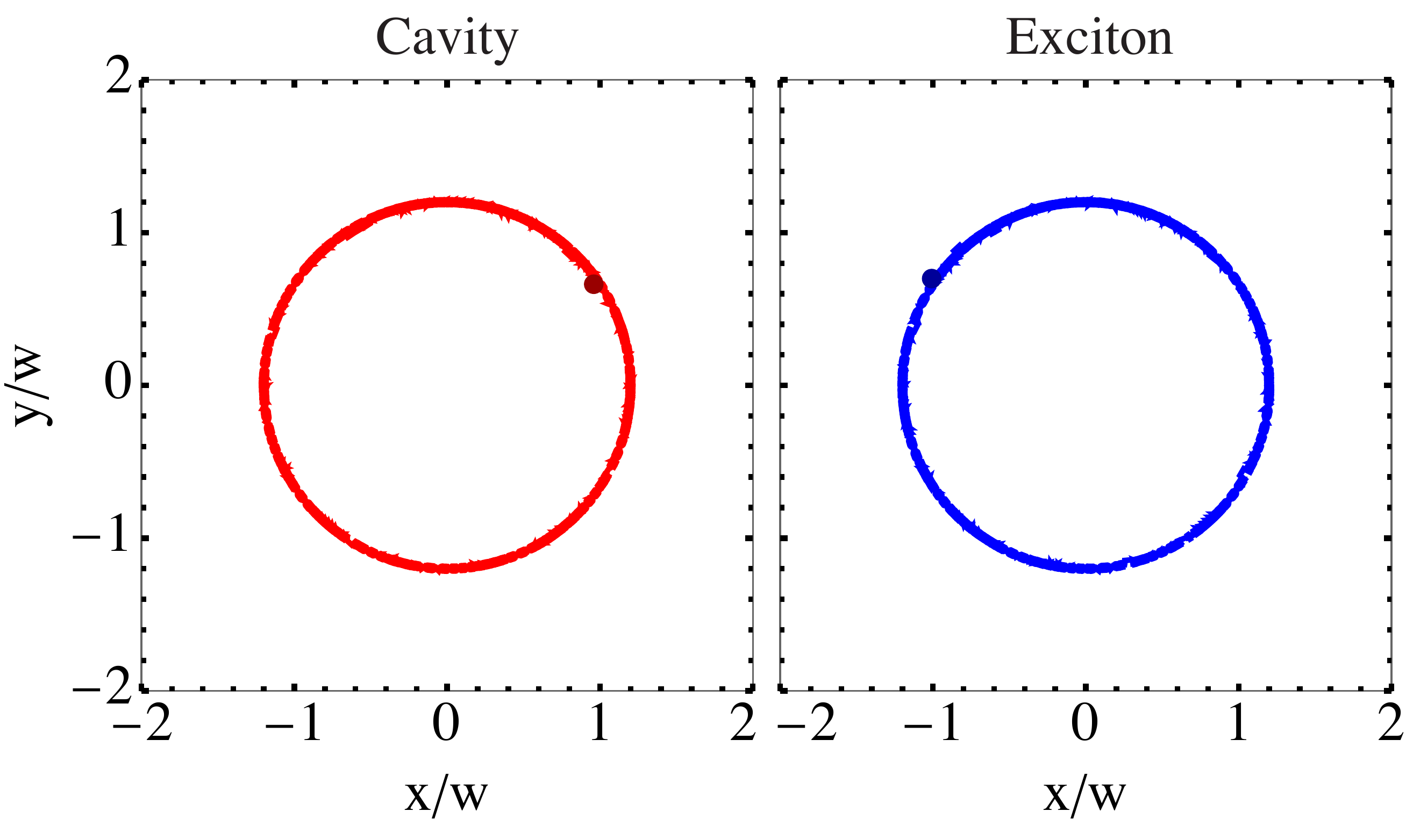}
\caption[Circular motion for high detunig]{Dynamics of vortex cores in a high detuning regime, where the vortex cores follow a circular trajectory. This scenario presents a fascinating parallel with the classical approach. Here, we use $\hbar\Omega = \qty{2.75}{\milli\electronvolt}$, $\hbar\Delta = \qty{100}{\milli\electronvolt}$ , $ \mathrm{w} = 25\, \mu\text{m}$, and $\mathrm{L}_{0}=3\sqrt{2}\mathrm{w}/5$. The initial position of the cores are $\mathrm{x}_{\mathrm{C}} = \mathrm{L}_{0}$, $\mathrm{y}_{\mathrm{C}} = \mathrm{L}_{0}$, $\mathrm{x}_{\mathrm{X}} = -\mathrm{L}_{0}$, and $\mathrm{y}_{\mathrm{X}} = -\mathrm{L}_{0}$.}
\label{fig:circtrajectories}
\end{figure}

On the other hand, when the system is considered in the high detuning regime, where $\Delta$ significantly exceeds the coupling constant $\Omega$, the vortex cores exhibit circular trajectories as is shown in Fig.~\ref{fig:circtrajectories}. This particular regime is distinguished by weak coupling, which implies a lack of strong correlations between the different components. Interestingly, this dynamics mimics the observed behavior of the vortex core in classical regimes. Such resemblance provides a theoretical link to the work of Rahmani et al. ~\cite{Rahmani2019}, who noted a similar pattern. In their framework, a mean-field approach demonstrated how the vortex cores followed a circular path under certain conditions. This similarity suggests that the high detuning regime shares some fundamental properties with classical states, particularly regarding the vortex core dynamics. In contrast with the simple nature of these classical-like vortex core trajectories, quantum systems demonstrate a richer and more intricate dynamic behavior as seen in Fig.~\ref{fig:trajectories}. 
In each case, we can conclude that detuning is a critical parameter that allows precise control over the trajectory's shape, which also depends critically on the relative phase introduced by the distance between the two vortex cores (photonic and excitonic). 

\begin{figure}[p]
\centering
\includegraphics[width=0.95\textwidth]{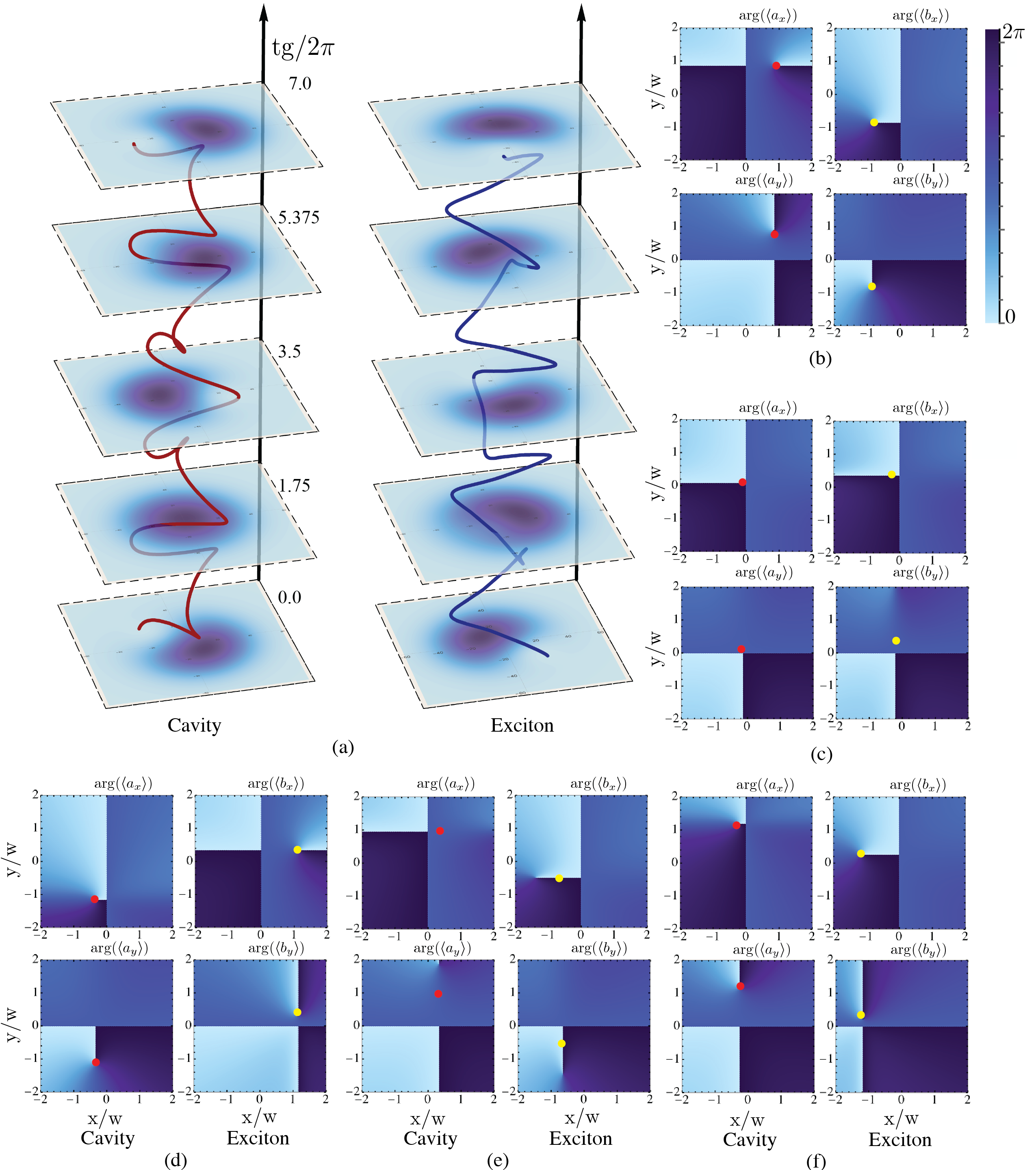}
\caption{Quantum dynamics of the vortex core for the initial quantum state. Panel (a) shows the trajectories of the core for the cavity photon (left) and exciton (right) fields.  Panels (b)-(f) depict the phase of the annihilation operator at each time of panel (a).  Here, we use $\hbar\Omega = \qty{2.75}{\milli\electronvolt}$, $\hbar\Delta = \qty{-1.5}{\milli\electronvolt}$ , $ \mathrm{w} = 25\, \mu\text{m}$, and $\mathrm{L}_{0}=3\sqrt{2}\mathrm{w}/5$. The initial position of the cores are $\mathrm{x}_{\mathrm{C}} = \mathrm{L}_{0}$, $\mathrm{y}_{\mathrm{C}} = \mathrm{L}_{0}$, $\mathrm{x}_{\mathrm{X}} = -\mathrm{L}_{0}$, and $\mathrm{y}_{\mathrm{X}} = -\mathrm{L}_{0}$.}
\label{fig:qdynamics}
\end{figure}

Fig.~\ref{fig:qdynamics} shows the time evolution of the vortex core under the initial conditions given in~(\ref{eq:incond}), which corresponds to the preparation of a vortex state by a Q-plate onto the vacuum photon field and the exciton field. Panel (a) depicts the vortex trajectory for the cavity photon and the exciton, respectively. This perspective eloquently shows how each component goes through an unconventional path that departs from the helical shape obtained for the semi-classical approach~\cite{Rahmani2019}. Additionally, on each inset plane, we have drawn the probability density together with the core position at times $\mathrm{tg}/2\pi=\{0,\allowbreak 7/4,\allowbreak 7/2,\allowbreak 43/8,\allowbreak 7\}$
(see the supplementary movies {\it QuantumTrajectory} and {\it QuantumTTail}~\cite{supvideos:2023} for a clear representation of the dynamics on the Riemann sphere, which will be described later in this section). 

The wave function's phase has an inevitable discontinuity of $2\pi$ in the presence of a vortex. Hence, it is an key theoretical and experimental criterion to characterize the actual formation of a core of rotation in a given system. In general, there is no wave function here; therefore, the phase of the annihilation operator benchmarks the main phenomena related to the phase in a pure state. The panels (b)-(f) show the argument (phase) of the mean value calculated for the field operator, i.e., the annihilation operators $\arg(\braket{a}), \arg(\braket{b})$.

\begin{figure}[ht]
\centering
\includegraphics[width=0.95\textwidth]{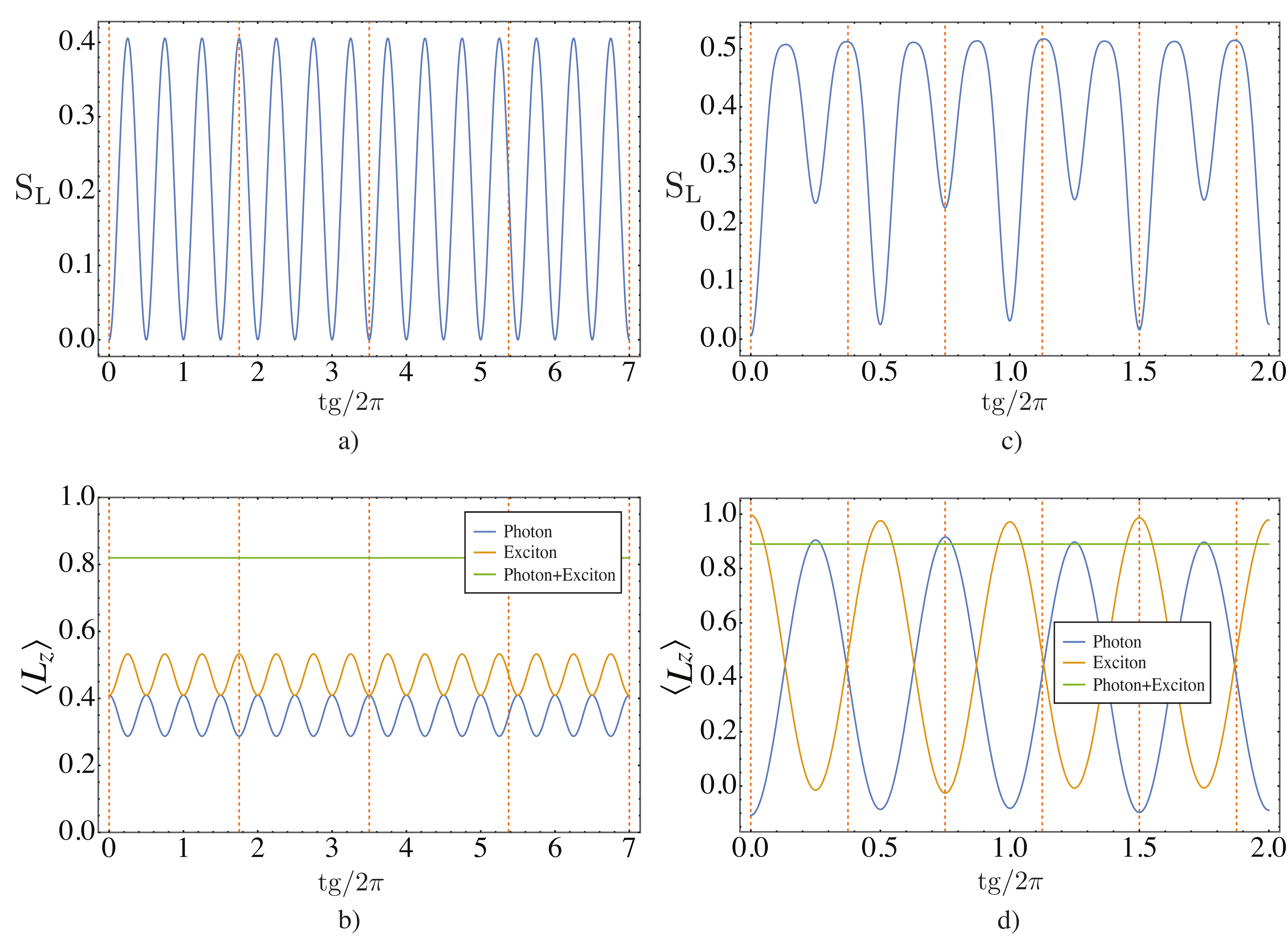}
\caption{Quantum vortex initial state: (a) Linear entropy measures the entanglement between photonic and excitonic components as a function of time. (b) Angular momentum ($z$ projection) for each field component and the whole system. Coherent vortex initial state: (c) Linear entropy  (d) Angular momentum ($z$ projection).}
    \label{fig:entquantumcohe}
\end{figure}

A red (yellow) point shows the vortex core position for the photonic (excitonic) component. For this initial condition, the panels show the phase $\arg({a_i})$ or $\arg{(b_i)}$ for the four fields involved. As expected, each field has a phase jump of $2\pi$ at the location of the vortex core. However, the core coincides with the phase jump vertex (the coordinate where the phase discontinuity has its origin) only when the state is completely disentangled, as reported in Fig.~\ref{fig:entquantumcohe}~a) where the relevant times are highlighted in each case with dashed vertical lines. At times $\mathrm{tg}/2\pi=\{0,\allowbreak 7/2,\allowbreak 7\}$ the linear entropy is identically zero, meaning there is no entanglement between the two components. On the other hand, at maximum entanglement ($\mathrm{tg}/2\pi=7/4$), it is impossible to identify the core with the phase jump vertex. However, the core remains located at the phase jump line. The same analysis is valid for other values of entanglement different from zero. The angular momentum is drawn in Fig.~\ref{fig:entquantumcohe} b); the initial condition for both components has no difference. However, the angular momentum difference is maximal when the entanglement has a higher value. In contrast, when the entanglement disappears, at the most distant turning points where the velocity diminishes, both values of angular momentum coincide. When there is maximal entanglement, the angular momentum of the cavity field is at its minimum, whereas the excitonic field reaches its maximum value. 

\begin{figure}[p]
\centering
    \includegraphics[width=0.95\textwidth]{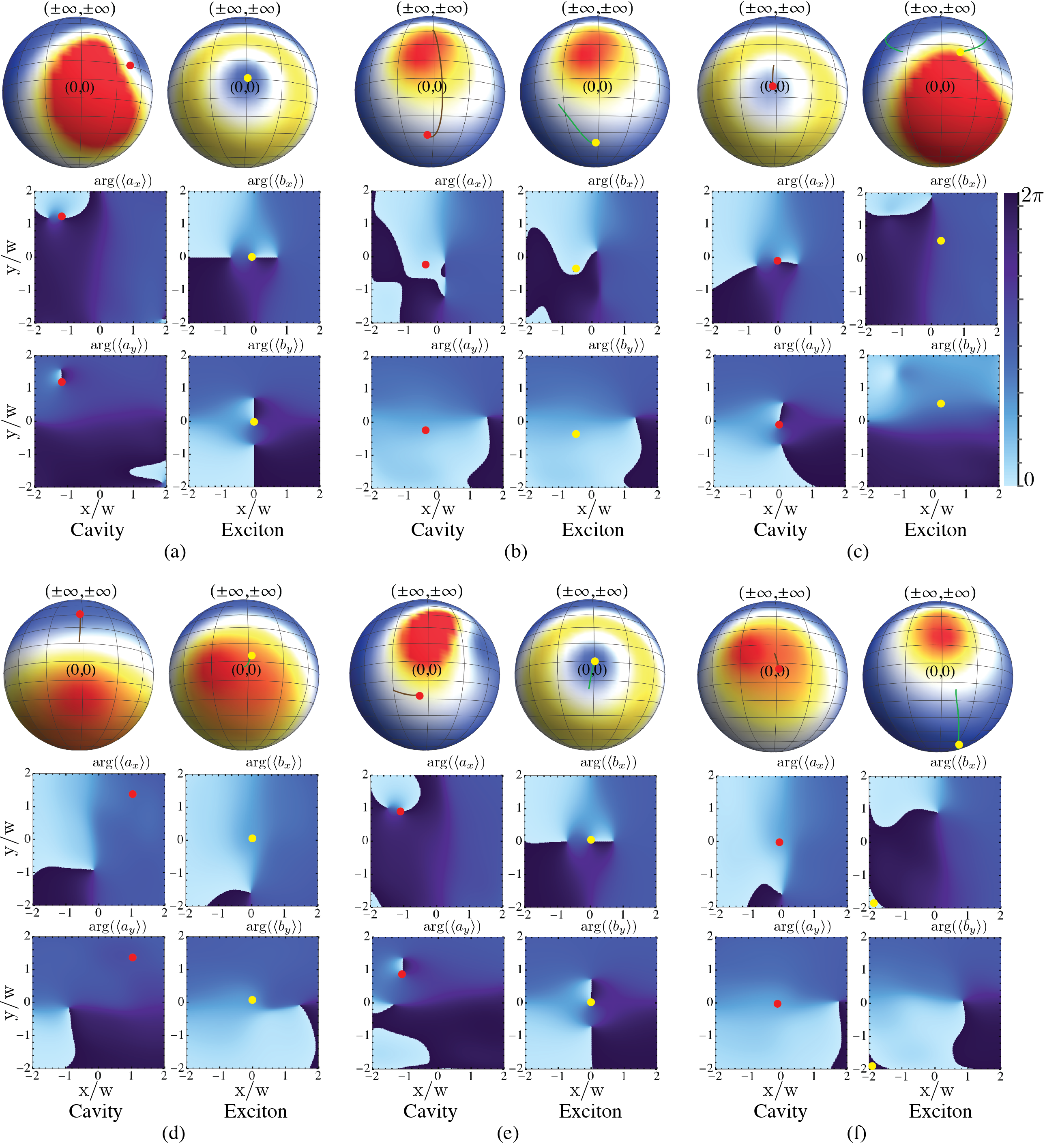}
\caption{Quantum dynamics of the vortex core for the coherent initial state. Each panel shows the trajectory of the vortex core, along with the probability density,  for the cavity photon (left) and exciton (right) fields over the surface of the Riemann sphere. The blue region represents a null density value, whereas the red represents its maximum value.  Additionally, each panel depicts the phase of the annihilation operators at times $\mathrm{tg}/2\pi$: (a) 
 $0$, (b) $3/8$, (c) $3/4$, (d) $9/8$, (e) $3/2$, (f) $15/8$.  Here, we use $\hbar\Omega = \qty{2.75}{\milli\electronvolt}$, $\hbar\Delta = \qty{-1.5}{\milli\electronvolt}$, $ \mathrm{w} = 25\, \mu\text{m}$, and $\mathrm{L}_{0}=6\mathrm{w}/5$. The initial position of the cores are $\mathrm{x_C} =-\mathrm{L}_{0}$, $\mathrm{y_C} = \mathrm{L}_{0}$, $\mathrm{x_X} = \qty{0}{\mu \textrm{m}}$, and $\mathrm{y_X} = \qty{0}{\mu \textrm{m}}$. The coherent states are $\ket{\alpha_x=(1/\sqrt{6})(-1+i)}$, $\ket{\alpha_y=(1/\sqrt{6})(1+i)}$, $\ket{\beta_x=0}$, and $\ket{\beta_y=0}$. The south pole of the Sphere is at the coordinate point $(-\sqrt{2}\mathrm{R},-\sqrt{2}\mathrm{R}$). This allows the $xy$ plane's origin to be projected onto the sphere equator. $\mathrm{R}$ is the sphere radius whose value can be arbitrarily chosen.} 
    \label{fig:cohdynamics}
\end{figure}

The dynamics of a coherent vortex introduced in the~\ref{app:cohIC}, is presented in Fig.~\ref{fig:cohdynamics}.  We depict the trajectory and the densities on the stereographic projection of the $xy$ plane over the surface of the Riemann sphere. In this representation the north pole corresponds to  $\widetilde{\infty}$, which symbolizes the confluence of all the infinities associated with the coordinate plane. Again, the core trajectories are far from trivial. Once the angular moment is transferred to the coherent states, they display a continuum modification of its densities in such a way that the core seems to be quickly transferred between both components (see also the supplementary movies \textit{CoherentTrajectory} and \textit{CoherentTail}~\cite{supvideos:2023} for the representation of the dynamics). This, in turn, can be interpreted as a quick and cyclical transference of angular momentum  between the two components as the interaction takes place.

The vortex location is represented on the sphere by a red (yellow) point for the cavity photon (exciton). Besides, attached to this point is a tail whose length is proportional to the core velocity: the longer the tail is, the faster the core goes.  The vortex core movement is slow when it is close to the origin and fast when expelled from the center. As we have done for the quantum initial condition case, the phase of the annihilation operator mean value is shown in each panel, along with the position of the vortex core at a given time: 
$\mathrm{tg}/2\pi=\{0,\allowbreak 3/8,\allowbreak 3/4,\allowbreak 9/8,\allowbreak 3/2,\allowbreak 15/8\}$. As before, we have chosen this set of times to graphically show specific levels of the photon-exciton entanglement in  Fig.~\ref{fig:entquantumcohe}~c). The relevant times are highlighted in each case with dashed vertical lines. At times $\mathrm{tg}/2\pi=\{0,\allowbreak 3/2\}$ the linear entropy becomes quite small, i.e., the state is almost disentangled  and the core nearly coincides with the phase jump vertex. When the entanglement is at its maximum, the vortex core location departs from the curves of phase jumps. In this case, the phase jumps have a more complex structure, displaying loops where the phase undergoes a $2\pi$ jump. Despite this, we can always detect a phase vertex within the graph.  On the other hand, the angular momentum is drawn in Fig.~\ref{fig:entquantumcohe}~d). In this case,  the initial condition for both components has the maximum difference. Then, the angular momentum difference is maximal when the entanglement vanishes. Otherwise, when it has its maximum value at a turning point, where the velocity increases briskly, the vortex core moves away just before reappearing.

\subsection{Experimental realization proposal}
The injection of quantum vortices in a semiconductor microcavity and the reconstruction of their trajectories compose two main experimental challenges: first, non-radiative absorption and scattering processes considerably reduce the system's transmission, limiting the detection capabilities of an actual setup. Second, the system's trajectory reconstruction needs to be done through a technique named ``off-axis digital holography'' that relies on the interference between the transmitted signal and a reference beam with homogeneous intensity and phase profiles ~\cite{Quinteiro2022,Donati2016}. This technique is widely used for ultrafast optical imaging by extracting the information encoded in the Fourier Transform of the interference pattern. However, the demonstration of a single photon implementation of such a technique has not been reported yet. It is important to emphasize that none of these challenges represents a fundamental limitation. Moreover, steps towards the imaging of polaritonic systems at the single particle regime have been reported in recent years~\cite{Suarez-Forero2020}, demonstrating that it is possible to overcome the non-radiative losses of the system to obtain space-resolved images of propagating single polaritons. Regarding holographic techniques, a practical implementation would imply having a phase-correlated reference of the injected single photons, which could be achieved with SPDC sources.  In this case, heralded measurements with spatial resolution would be necessary. In this regard, there is a reported demonstration of the use of photon pairs to inject single excitations in a microcavity-quantum well system and a successive two qubits quantum tomography to reconstruct the state~\cite{Cuevas2018}.

With these considerations in mind, we can now propose an experimental implementation to reconstruct the dynamics of a single polariton vortex. In this case, a pulsed laser would be split in two; one branch, with an expanded spatial profile and homogeneous phase, would go through a delay line and reach the detection, which could be a CCD camera; this first beam would be used as a reference. The other portion of the laser pulse would go over two optical parametric oscillators. The first process would double the frequency to obtain a higher energy pulse that would be directed again into a nonlinear crystal to generate photon pairs. One of the generated photons would pass through a device to imprint the vortex phase profile (namely a spatial light modulator or a Q-plate), and then be directed into the cavity. After its injection, propagation, and re-emission, the photon would be recombined with the reference beam, and, since they have respectively locked phases, a fast Fourier transform would allow the extraction of the information at simultaneity. Therefore, upon repeating the experiment for different time delays between reference and signal, the dynamics (trajectory) of the single polariton vortex could be reconstructed.

\section{Conclusions}

A completely quantum approach for the polariton vortex dynamics has been provided in a suitable platform for describing vorticity beyond the usual theoretical formulations.

We have shown that, for different analyzed scenarios, it can be ascertained that the detuning serves as a pivotal parameter permitting meticulous control over the shape of the trajectory. These trajectories are further critically influenced by the relative phase determined by the distance between the two vortex cores, i.e., photonic and excitonic ones. 
The exploration of polariton vortex dynamics through a quantum approach unveils a remarkable platform for investigating vorticity that extends well beyond conventional experimental applications. With quantum dynamics in focus, there emerges a set of quite interesting trajectories of the vortex core, largely attributed to the genuine quantum superposition interplay between light and matter.  As a result, the vortex core trajectories differ from the common circular paths, which establishes a qualitative criterion to differentiate the behavior at the edge between the quantum and the common macroscopic well-studied polariton vortex dynamics.  This outcome suggests that the entanglement between polariton components plays a significant role in shaping the overall dynamics of the system. The insights gained from this quantum approach open up new avenues for exploring quantum systems, promising advancements in both theoretical understanding and practical applications.

Finally, we have made a feasible experimental proposal that can serve as a guide to implement our theoretical model in order to observe the described phenomena formulated in this study.

\section{Acknowledgments}

We gratefully acknowledge funding by Universidad Nacional de Colombia under the project ``Hermes 57522: Centro de excelencia en tecnologías cuánticas y sus aplicaciones a metrología''. J.P.R.C. gratefully acknowledges support from the “Beca de Doctorados Nacionales de MINCIENCIAS 785”

\bibliographystyle{unsrtnat}
\bibliography{Ref}

\begin{thebibliography}{48}
\providecommand{\natexlab}[1]{#1}
\providecommand{\url}[1]{\texttt{#1}}
\expandafter\ifx\csname urlstyle\endcsname\relax
  \providecommand{\doi}[1]{doi: #1}\else
  \providecommand{\doi}{doi: \begingroup \urlstyle{rm}\Url}\fi

\bibitem[Pines and Nozi\`eres(1966)]{Pines1966}
David Pines and Philippe Nozi\`eres.
\newblock \emph{The Theory of Quantum Liquids}.
\newblock CRC Press, 1966.
\newblock \doi{10.4324/9780429492662}.

\bibitem[Leggett(2004)]{Leggett2004}
Anthony~J. Leggett.
\newblock Nobel lecture: Superfluid $^{3}\mathrm{He}$: the early days as seen
  by a theorist.
\newblock \emph{Rev. Mod. Phys.}, 76:\penalty0 999--1011, Dec 2004.
\newblock \doi{10.1103/RevModPhys.76.999}.

\bibitem[Giorgini et~al.(2008)Giorgini, Pitaevskii, and
  Stringari]{Giorgini2008}
Stefano Giorgini, Lev~P. Pitaevskii, and Sandro Stringari.
\newblock Theory of ultracold atomic {F}ermi gases.
\newblock \emph{Rev. Mod. Phys.}, 80:\penalty0 1215--1274, Oct 2008.
\newblock \doi{10.1103/RevModPhys.80.1215}.

\bibitem[Carusotto and Ciuti(2013)]{Carusotto2013}
Iacopo Carusotto and Cristiano Ciuti.
\newblock Quantum fluids of light.
\newblock \emph{Rev. Mod. Phys.}, 85:\penalty0 299--366, Feb 2013.
\newblock \doi{10.1103/RevModPhys.85.299}.

\bibitem[Bewley et~al.(2008)Bewley, Paoletti, Sreenivasan, and
  Lathrop]{Bewley08}
Gregory~P. Bewley, Matthew~S. Paoletti, Katepalli~R. Sreenivasan, and Daniel~P.
  Lathrop.
\newblock Characterization of reconnecting vortices in superfluid helium.
\newblock \emph{Proceedings of the National Academy of Sciences}, 105\penalty0
  (37):\penalty0 13707--13710, 2008.
\newblock \doi{10.1073/pnas.0806002105}.

\bibitem[Henn et~al.(2009)Henn, Seman, Roati, Magalh\~aes, and Bagnato]{Henn09}
E.~A.~L. Henn, J.~A. Seman, G.~Roati, K.~M.~F. Magalh\~aes, and V.~S. Bagnato.
\newblock Emergence of turbulence in an oscillating {B}ose-{E}instein
  condensate.
\newblock \emph{Phys. Rev. Lett.}, 103:\penalty0 045301, Jul 2009.
\newblock \doi{10.1103/PhysRevLett.103.045301}.

\bibitem[Dominici et~al.(2018)Dominici, Carretero-Gonz{\'a}lez, Gianfrate,
  Cuevas-Maraver, Rodrigues, Frantzeskakis, Lerario, Ballarini, De~Giorgi,
  Gigli, Kevrekidis, and Sanvitto]{Dominici18}
Lorenzo Dominici, Ricardo Carretero-Gonz{\'a}lez, Antonio Gianfrate, Jes{\'u}s
  Cuevas-Maraver, Augusto~S. Rodrigues, Dimitri~J. Frantzeskakis, Giovanni
  Lerario, Dario Ballarini, Milena De~Giorgi, Giuseppe Gigli, Panayotis~G.
  Kevrekidis, and Daniele Sanvitto.
\newblock Interactions and scattering of quantum vortices in a polariton fluid.
\newblock \emph{Nature Communications}, 9\penalty0 (1):\penalty0 1467, 2018.
\newblock \doi{10.1038/s41467-018-03736-5}.

\bibitem[Zhao et~al.(2017)Zhao, Misko, Tempere, and Nori]{Zhao2017}
H.~J. Zhao, V.~R. Misko, J.~Tempere, and F.~Nori.
\newblock Pattern formation in vortex matter with pinning and frustrated
  intervortex interactions.
\newblock \emph{Phys. Rev. B}, 95:\penalty0 104519, Mar 2017.
\newblock \doi{10.1103/PhysRevB.95.104519}.

\bibitem[Feynman(1955)]{feynman55a}
R.P. Feynman.
\newblock Chapter {II}. {A}pplication of quantum mechanics to liquid helium.
\newblock In C.J. Gorter, editor, \emph{Progress in Low Temperature Physics},
  volume~1, pages 17--53. Elsevier, 1955.
\newblock \doi{10.1016/S0079-6417(08)60077-3}.

\bibitem[Alperin and Berloff(2021)]{Alperin21}
Samuel~N. Alperin and Natalia~G. Berloff.
\newblock Multiply charged vortex states of polariton condensates.
\newblock \emph{Optica}, 8\penalty0 (3):\penalty0 301, Mar 2021.
\newblock \doi{10.1364/OPTICA.418377}.

\bibitem[Shin et~al.(2004)Shin, Saba, Vengalattore, Pasquini, Sanner,
  Leanhardt, Prentiss, Pritchard, and Ketterle]{Shin04}
Y.~Shin, M.~Saba, M.~Vengalattore, T.~A. Pasquini, C.~Sanner, A.~E. Leanhardt,
  M.~Prentiss, D.~E. Pritchard, and W.~Ketterle.
\newblock Dynamical instability of a doubly quantized vortex in a
  {B}ose-{E}instein condensate.
\newblock \emph{Phys. Rev. Lett.}, 93:\penalty0 160406, Oct 2004.
\newblock \doi{10.1103/PhysRevLett.93.160406}.

\bibitem[Engels et~al.(2003)Engels, Coddington, Haljan, Schweikhard, and
  Cornell]{Engels03}
P.~Engels, I.~Coddington, P.~C. Haljan, V.~Schweikhard, and E.~A. Cornell.
\newblock Observation of long-lived vortex aggregates in rapidly rotating
  {B}ose-{E}instein condensates.
\newblock \emph{Phys. Rev. Lett.}, 90:\penalty0 170405, May 2003.
\newblock \doi{10.1103/PhysRevLett.90.170405}.

\bibitem[Sasaki et~al.(2010)Sasaki, Suzuki, and Saito]{Sasaki10}
Kazuki Sasaki, Naoya Suzuki, and Hiroki Saito.
\newblock B\'enard--von {K}\'arm\'an vortex street in a {B}ose-{E}instein
  condensate.
\newblock \emph{Phys. Rev. Lett.}, 104:\penalty0 150404, Apr 2010.
\newblock \doi{10.1103/PhysRevLett.104.150404}.

\bibitem[Dominici et~al.(2015)Dominici, Dagvadorj, Fellows, Ballarini, {De
  Giorgi}, Marchetti, Piccirillo, Marrucci, Bramati, Gigli, Szyma{\~{n}}ska,
  and Sanvitto]{Dominici2015}
Lorenzo Dominici, Galbadrakh Dagvadorj, Jonathan~M. Fellows, Dario Ballarini,
  Milena {De Giorgi}, Francesca~M. Marchetti, Bruno Piccirillo, Lorenzo
  Marrucci, Alberto Bramati, Giuseppe Gigli, Marzena~H. Szyma{\~{n}}ska, and
  Daniele Sanvitto.
\newblock {Vortex and half-vortex dynamics in a nonlinear spinor quantum
  fluid}.
\newblock \emph{Sci. Adv.}, 1\penalty0 (11):\penalty0 1--10, 2015.
\newblock \doi{10.1126/sciadv.1500807}.

\bibitem[Dominici et~al.(2021)Dominici, Colas, Gianfrate, Rahmani, Ardizzone,
  Ballarini, {De Giorgi}, Gigli, Laussy, Sanvitto, and Voronova]{Dominici2021a}
Lorenzo Dominici, David Colas, Antonio Gianfrate, Amir Rahmani, Vincenzo
  Ardizzone, Dario Ballarini, Milena {De Giorgi}, Giuseppe Gigli, Fabrice~P.
  Laussy, Daniele Sanvitto, and Nina Voronova.
\newblock {Full-Bloch beams and ultrafast {R}abi-rotating vortices}.
\newblock \emph{Phys. Rev. Res.}, 3\penalty0 (1):\penalty0 013007, 2021.
\newblock \doi{10.1103/physrevresearch.3.013007}.

\bibitem[Fisher et~al.(1996)Fisher, Afshar, Skolnick, Whittaker, and
  Roberts]{fisher96a}
T.~A. Fisher, A.~M. Afshar, M.~S. Skolnick, D.~M. Whittaker, and J.~S. Roberts.
\newblock Vacuum {R}abi coupling enhancement and {Zeeman} splitting in
  semiconductor quantum microcavity structures in a high magnetic field.
\newblock \emph{Phys. Rev. B}, 53:\penalty0 R10469, 1996.
\newblock \doi{10.1103/physrevb.53.r10469}.

\bibitem[Fetter and Svidzinsky(2001)]{fetter01a}
A.L. Fetter and A.A. Svidzinsky.
\newblock Vortices in a trapped dilute {Bose}--{Einstein} condensate.
\newblock \emph{Journal of Physics: Condensed Matter}, 13\penalty0
  (12):\penalty0 R135, 2001.
\newblock \doi{10.1088/0953-8984/13/12/201}.

\bibitem[Li et~al.(2018)Li, Sampuli, Song, Xia, and Ding]{Li18}
Chuang Li, Elijah~M Sampuli, Jie Song, Yan Xia, and Weiqiang Ding.
\newblock One-step engineering many-atom {NOON} state.
\newblock \emph{New Journal of Physics}, 20\penalty0 (9):\penalty0 093019,
  2018.
\newblock \doi{10.1088/1367-2630/aadf9e}.

\bibitem[Su\'arez-Forero et~al.(2016)Su\'arez-Forero, Cipagauta, Vinck-Posada,
  Fonseca~Romero, Rodr\'{\i}guez, and Ballarini]{Suarez16}
D.~G. Su\'arez-Forero, G.~Cipagauta, H.~Vinck-Posada, K.~M. Fonseca~Romero,
  B.~A. Rodr\'{\i}guez, and D.~Ballarini.
\newblock Entanglement properties of quantum polaritons.
\newblock \emph{Phys. Rev. B}, 93:\penalty0 205302, May 2016.
\newblock \doi{10.1103/PhysRevB.93.205302}.

\bibitem[Deng et~al.(2010)Deng, Haug, and Yamamoto]{deng10a}
Hui Deng, Hartmut Haug, and Yoshihisa Yamamoto.
\newblock Exciton-polariton {B}ose-{E}instein condensation.
\newblock \emph{Rev. Mod. Phys.}, 82:\penalty0 1489, 2010.
\newblock \doi{10.1103/RevModPhys.82.1489}.

\bibitem[Amo et~al.(2009)Amo, Sanvitto, Laussy, Ballarini, Valle, Martin,
  Lema{\^\i}tre, Bloch, Krizhanovskii, Skolnick, Tejedor, and
  Vi{\~n}a]{Amo2009}
A.~Amo, D.~Sanvitto, F.~P. Laussy, D.~Ballarini, E.~del Valle, M.~D. Martin,
  A.~Lema{\^\i}tre, J.~Bloch, D.~N. Krizhanovskii, M.~S. Skolnick, C.~Tejedor,
  and L.~Vi{\~n}a.
\newblock Collective fluid dynamics of a polariton condensate in a
  semiconductor microcavity.
\newblock \emph{Nature}, 457\penalty0 (7227):\penalty0 291, 2009.
\newblock \doi{10.1038/nature07640}.

\bibitem[Dominici et~al.(2014)Dominici, Colas, Donati, Restrepo~Cuartas,
  De~Giorgi, Ballarini, Guirales, L\'opez Carre\~no, Bramati, Gigli, del Valle,
  Laussy, and Sanvitto]{Dominici2014}
L.~Dominici, D.~Colas, S.~Donati, J.~P. Restrepo~Cuartas, M.~De~Giorgi,
  D.~Ballarini, G.~Guirales, J.~C. L\'opez Carre\~no, A.~Bramati, G.~Gigli,
  E.~del Valle, F.~P. Laussy, and D.~Sanvitto.
\newblock Ultrafast control and {R}abi oscillations of polaritons.
\newblock \emph{Phys. Rev. Lett.}, 113:\penalty0 226401, Nov 2014.
\newblock \doi{10.1103/PhysRevLett.113.226401}.

\bibitem[Marie et~al.(1999)Marie, Renucci, Dubourg, Amand, Le~Jeune, Barrau,
  Bloch, and Planel]{Marie1999}
X.~Marie, P.~Renucci, S.~Dubourg, T.~Amand, P.~Le~Jeune, J.~Barrau, J.~Bloch,
  and R.~Planel.
\newblock Coherent control of exciton polaritons in a semiconductor
  microcavity.
\newblock \emph{Phys. Rev. B}, 59:\penalty0 R2494--R2497, Jan 1999.
\newblock \doi{10.1103/PhysRevB.59.R2494}.

\bibitem[Quinteiro~Rosen et~al.(2022)Quinteiro~Rosen, Tamborenea, and
  Kuhn]{Quinteiro2022}
Guillermo~F. Quinteiro~Rosen, Pablo~I. Tamborenea, and Tilmann Kuhn.
\newblock Interplay between optical vortices and condensed matter.
\newblock \emph{Rev. Mod. Phys.}, 94:\penalty0 035003, Aug 2022.
\newblock \doi{10.1103/RevModPhys.94.035003}.

\bibitem[Gnusov et~al.(2023)Gnusov, Harrison, Alyatkin, Sitnik, T{\"o}pfer,
  Sigurdsson, and Lagoudakis]{Gnusov2023}
Ivan Gnusov, Stella Harrison, Sergey Alyatkin, Kirill Sitnik, Julian
  T{\"o}pfer, Helgi Sigurdsson, and Pavlos Lagoudakis.
\newblock Quantum vortex formation in the ``rotating bucket'' experiment with
  polariton condensates.
\newblock \emph{Science Advances}, 9\penalty0 (4):\penalty0 eadd1299, 2023.
\newblock \doi{10.1126/sciadv.add1299}.

\bibitem[Wang et~al.(2022)Wang, Peng, Xu, Feng, Huang, Wu, Liew, and
  Xiong]{Wang2022}
Jun Wang, Yutian Peng, Huawen Xu, Jiangang Feng, Yuqing Huang, Jinqi Wu,
  Timothy C~H Liew, and Qihua Xiong.
\newblock {Controllable vortex lasing arrays in a geometrically frustrated
  exciton–polariton lattice at room temperature}.
\newblock \emph{National Science Review}, 10\penalty0 (1):\penalty0 nwac096, 05
  2022.
\newblock \doi{10.1093/nsr/nwac096}.

\bibitem[Lagoudakis et~al.(2008)Lagoudakis, Wouters, Richard, Baas, Carusotto,
  Andr{\'{e}}, Dang, and Deveaud-Pl{\'{e}}dran]{Lagoudakis2008}
K.~G. Lagoudakis, M.~Wouters, M.~Richard, A.~Baas, I.~Carusotto,
  R.~Andr{\'{e}}, Le~Si Dang, and B.~Deveaud-Pl{\'{e}}dran.
\newblock {Quantized vortices in an exciton-polariton condensate}.
\newblock \emph{Nat. Phys.}, 4\penalty0 (9):\penalty0 706, 2008.
\newblock \doi{10.1038/nphys1051}.

\bibitem[Lagoudakis et~al.(2009)Lagoudakis, Ostatnick{\'y}, Kavokin, Rubo,
  Andr{\'e}, and Deveaud-Pl{\'e}dran]{Lagoudakis2009}
K.~G. Lagoudakis, T.~Ostatnick{\'y}, A.~V. Kavokin, Y.~G. Rubo, R.~Andr{\'e},
  and B.~Deveaud-Pl{\'e}dran.
\newblock Observation of half-quantum vortices in an exciton-polariton
  condensate.
\newblock \emph{Science}, 326\penalty0 (5955):\penalty0 974--976, 2009.
\newblock \doi{10.1126/science.1177980}.

\bibitem[Redondo et~al.(2022)Redondo, Schneider, Klembt, Höfling, Tarucha, and
  Fraser]{Redondo2022}
Yago del Valle~Inclan Redondo, Christian Schneider, Sebastian Klembt, Sven
  Höfling, Seigo Tarucha, and Michael~D. Fraser.
\newblock Optically driven rotation of exciton-polariton condensates.
\newblock \emph{arXiv}, 2022.
\newblock \doi{10.48550/ARXIV.2209.01904}.

\bibitem[Sitnik et~al.(2022)Sitnik, Alyatkin, T\"opfer, Gnusov, Cookson,
  Sigurdsson, and Lagoudakis]{Sitnik2022}
Kirill~A. Sitnik, Sergey Alyatkin, Julian~D. T\"opfer, Ivan Gnusov, Tamsin
  Cookson, Helgi Sigurdsson, and Pavlos~G. Lagoudakis.
\newblock Spontaneous formation of time-periodic vortex cluster in nonlinear
  fluids of light.
\newblock \emph{Phys. Rev. Lett.}, 128:\penalty0 237402, Jun 2022.
\newblock \doi{10.1103/PhysRevLett.128.237402}.

\bibitem[Gao et~al.(2022)Gao, Hu, Schumacher, and Ma]{Xinghui2022}
Xinghui Gao, Wei Hu, Stefan Schumacher, and Xuekai Ma.
\newblock Unidirectional vortex waveguides and multistable vortex pairs in
  polariton condensates.
\newblock \emph{Opt. Lett.}, 47\penalty0 (13):\penalty0 3235--3238, Jul 2022.
\newblock \doi{10.1364/OL.457724}.

\bibitem[Dominici et~al.(2022)Dominici, Voronova, Rahmani, Colas, Ballarini,
  De~Giorgi, Gigli, Laussy, and Sanvitto]{Dominici2022}
Lorenzo Dominici, Nina Voronova, Amir Rahmani, David Colas, Dario Ballarini,
  Milena De~Giorgi, Giuseppe Gigli, Fabrice~P. Laussy, and Daniele Sanvitto.
\newblock Coupled quantum vortex kinematics and berry curvature in real space.
\newblock \emph{arXiv}, 2022.
\newblock \doi{10.48550/ARXIV.2202.13210}.

\bibitem[Horodecki et~al.(2009)Horodecki, Horodecki, Horodecki, and
  Horodecki]{Horodecki2009}
Ryszard Horodecki, Pawe\l{} Horodecki, Micha\l{} Horodecki, and Karol
  Horodecki.
\newblock Quantum entanglement.
\newblock \emph{Rev. Mod. Phys.}, 81:\penalty0 865--942, Jun 2009.
\newblock \doi{10.1103/RevModPhys.81.865}.

\bibitem[Ram{\'\i}rez-Mu{\~n}oz et~al.(2018)Ram{\'\i}rez-Mu{\~n}oz,
  Restrepo~Cuartas, and Vinck-Posada]{Ramirez2018}
J.~E. Ram{\'\i}rez-Mu{\~n}oz, J.P. Restrepo~Cuartas, and Herbert Vinck-Posada.
\newblock Quantum correlations between two cavity qed systems coupled by a
  mechanical resonator.
\newblock \emph{The European Physical Journal B}, 91\penalty0 (11):\penalty0
  268, 2018.
\newblock \doi{10.1140/epjb/e2018-90438-4}.

\bibitem[Fraser et~al.(2009)Fraser, Roumpos, and Yamamoto]{Fraser2009a}
M~D Fraser, G~Roumpos, and Y~Yamamoto.
\newblock Vortex{\textendash}antivortex pair dynamics in an
  exciton{\textendash}polariton condensate.
\newblock \emph{New Journal of Physics}, 11\penalty0 (11):\penalty0 113048, nov
  2009.
\newblock \doi{10.1088/1367-2630/11/11/113048}.

\bibitem[Ma et~al.(2020)Ma, Kartashov, Gao, Torner, and
  Schumacher]{Xuekai2020a}
Xuekai Ma, Yaroslav~V. Kartashov, Tingge Gao, Lluis Torner, and Stefan
  Schumacher.
\newblock Spiraling vortices in exciton-polariton condensates.
\newblock \emph{Phys. Rev. B}, 102:\penalty0 045309, Jul 2020.
\newblock \doi{10.1103/PhysRevB.102.045309}.

\bibitem[Rahmani and Dominici(2019)]{Rahmani2019}
Amir Rahmani and Lorenzo Dominici.
\newblock {Detuning control of {R}abi vortex oscillations in light-matter
  coupling}.
\newblock \emph{Phys. Rev. B}, 100\penalty0 (9):\penalty0 094310, 2019.
\newblock \doi{10.1103/PhysRevB.100.094310}.

\bibitem[Petrov et~al.(2022)Petrov, Sokolenko, Kulya, Gorodetsky, and
  Chernykh]{Petrov2022}
Nikolay~V. Petrov, Bogdan Sokolenko, Maksim~S. Kulya, Andrei Gorodetsky, and
  Aleksey~V. Chernykh.
\newblock Design of broadband terahertz vector and vortex beams: I. review of
  materials and components.
\newblock \emph{Light: Advanced Manufacturing}, 3\penalty0 (43), 2022.
\newblock \doi{10.37188/lam.2022.043}.

\bibitem[Vinck-Posada et~al.(2007)Vinck-Posada, Rodriguez, Guimaraes, Cabo, and
  Gonzalez]{Vinck-Posada2007}
Herbert Vinck-Posada, Boris~A. Rodriguez, P.~S.~S. Guimaraes, Alejandro Cabo,
  and Augusto Gonzalez.
\newblock Photon emission as a source of coherent behavior of polaritons.
\newblock \emph{Phys. Rev. Lett.}, 98:\penalty0 167405, 2007.
\newblock \doi{10.1103/PhysRevLett.98.167405}.

\bibitem[Tabata et~al.(2007)Tabata, Martins, Oliveira, Lamas, Duarte, da~Silva,
  and Gusev]{Tabata2007}
A.~Tabata, M.~R. Martins, J.~B.~B. Oliveira, T.~E. Lamas, C.~A. Duarte,
  E.~C.~F. da~Silva, and G.~M. Gusev.
\newblock Many-body effects in wide parabolic {AlGaAs} quantum wells.
\newblock \emph{Journal of Applied Physics}, 102\penalty0 (9):\penalty0 093715,
  2007.
\newblock \doi{10.1063/1.2809418}.

\bibitem[Tabata et~al.(2010)Tabata, Oliveira, da~Silva, Lamas, Duarte, and
  Gusev]{Tabata2010}
A~Tabata, J~B~B Oliveira, E~C~F da~Silva, T~E Lamas, C~A Duarte, and G~M Gusev.
\newblock Excitons in undoped {AlGaAs}/{GaAs} wide parabolic quantum wells.
\newblock \emph{Journal of Physics: Conference Series}, 210:\penalty0 012052,
  2010.
\newblock \doi{10.1088/1742-6596/210/1/012052}.

\bibitem[Gusev et~al.(2002)Gusev, Quivy, Lamas, Leite, Bakarov, Toropov,
  Estibals, and Portal]{Gusev2002}
G.~M. Gusev, A.~A. Quivy, T.~E. Lamas, J.~R. Leite, A.~K. Bakarov, A.~I.
  Toropov, O.~Estibals, and J.~C. Portal.
\newblock Magnetotransport of a quasi-three-dimensional electron gas in the
  lowest landau level.
\newblock \emph{Phys. Rev. B}, 65:\penalty0 205316, May 2002.
\newblock \doi{10.1103/PhysRevB.65.205316}.

\bibitem[Kasprzak et~al.(2010)Kasprzak, Reitzenstein, Muljarov, Kistner,
  Schneider, Strauss, H{\"o}fling, Forchel, and Langbein]{Kasprzak2010}
J.~Kasprzak, S.~Reitzenstein, E.~A. Muljarov, C.~Kistner, C.~Schneider,
  M.~Strauss, S.~H{\"o}fling, A.~Forchel, and W.~Langbein.
\newblock Up on the jaynes--cummings ladder of a quantum-dot/microcavity
  system.
\newblock \emph{Nature Materials}, 9\penalty0 (4):\penalty0 304, 2010.
\newblock \doi{10.1038/nmat2717}.

\bibitem[Kavokin et~al.(2017)Kavokin, Baumberg, Malpuech, and
  Laussy]{Kavokin2017}
Alexey~V. Kavokin, Jeremy~J. Baumberg, Guillaume Malpuech, and Fabrice~P.
  Laussy.
\newblock \emph{Microcavities (2nd edn)}.
\newblock Oxford University Press, 2017.
\newblock \doi{10.1093/oso/9780198782995.001.0001}.

\bibitem[sup()]{supvideos:2023}
See the Supplementary Movies files: QuantumTrajectory for the video of the
  quantum vortex dynamics related to Fig.~\ref{fig:qdynamics} and QuantumTTail
  for the video of the same dynamics depicting only the tail (velocity). The
  movie files: CoherentTrajectory and CoherentTTail are respectively the same
  supplementary now to Fig.~\ref{fig:cohdynamics}.

\bibitem[Donati et~al.(2016)Donati, Dominici, Dagvadorj, Ballarini, {De
  Giorgi}, Bramati, Gigli, Rubo, Szyma{\'{n}}ska, and Sanvitto]{Donati2016}
Stefano Donati, Lorenzo Dominici, Galbadrakh Dagvadorj, Dario Ballarini, Milena
  {De Giorgi}, Alberto Bramati, Giuseppe Gigli, Yuri~G. Rubo, Marzena~Hanna
  Szyma{\'{n}}ska, and Daniele Sanvitto.
\newblock {Twist of generalized skyrmions and spin vortices in a polariton
  superfluid}.
\newblock \emph{Proceedings of the National Academy of Sciences of the United
  States of America}, 113\penalty0 (52):\penalty0 14926--14931, dec 2016.
\newblock \doi{10.1073/pnas.1610123114}.

\bibitem[Su\'arez-Forero et~al.(2020)Su\'arez-Forero, Ardizzone, {Covre da
  Silva}, Reindl, Fieramosca, Polimeno, Giorgi, Dominici, Pfeiffer, Gigli,
  Ballarini, Laussy, Rastelli, and Sanvitto]{Suarez-Forero2020}
Daniel~Gustavo Su\'arez-Forero, Vincenzo Ardizzone, Saimon~Filipe {Covre da
  Silva}, Marcus Reindl, Antonio Fieramosca, Laura Polimeno, Milena~De Giorgi,
  Lorenzo Dominici, Loren~N. Pfeiffer, Giuseppe Gigli, Dario Ballarini, Fabrice
  Laussy, Armando Rastelli, and Daniele Sanvitto.
\newblock Quantum hydrodynamics of a single particle.
\newblock \emph{Light: Science \& Applications}, 9\penalty0 (1):\penalty0 85,
  dec 2020.
\newblock ISSN 2047-7538.
\newblock \doi{10.1038/s41377-020-0324-x}.

\bibitem[Cuevas et~al.(2018)Cuevas, Carre{\~{n}}o, Silva, {De Giorgi},
  Su{\'{a}}rez-Forero, Mu{\~{n}}oz, Fieramosca, Cardano, Marrucci, Tasco,
  Biasiol, {Del Valle}, Dominici, Ballarini, Gigli, Mataloni, Laussy,
  Sciarrino, and Sanvitto]{Cuevas2018}
{\'{A}}lvaro Cuevas, Juan Camilo~L{\'{o}}pez Carre{\~{n}}o, Blanca Silva,
  Milena {De Giorgi}, Daniel~G. Su{\'{a}}rez-Forero, Carlos~S{\'{a}}nchez
  Mu{\~{n}}oz, Antonio Fieramosca, Filippo Cardano, Lorenzo Marrucci,
  Vittorianna Tasco, Giorgio Biasiol, Elena {Del Valle}, Lorenzo Dominici,
  Dario Ballarini, Giuseppe Gigli, Paolo Mataloni, Fabrice~P. Laussy, Fabio
  Sciarrino, and Daniele Sanvitto.
\newblock {First observation of the quantized exciton-polariton field and
  effect of interactions on a single polariton}.
\newblock \emph{Science Advances}, 4\penalty0 (4):\penalty0 eaao6814, apr 2018.
\newblock \doi{10.1126/sciadv.aao6814}.

\end{thebibliography}

\appendix

\section{Coherent vortex}\label{app:cohIC}

In order to use a \textit{coherent vortex} as an initial condition for the dynamics,  we start from the coordinate representation of a coherent state. It is spanned by the Hermite basis times the core function $\mathrm{Z_k}(x,y,0)$ with $\mathrm{k=\{C,X\}}$

\begin{equation}
\psi^\mathrm{Z_k}_{\alpha_x,\alpha_y}(0)= \psi_{\alpha_x,\alpha_y}\mathrm{Z_k}(x,y,0).
\end{equation}

From the definition of a coherent state in the Fock representation

\begin{equation}
\ket{\alpha}= \exp\left(-\frac{|\alpha|^2}{2}\right)\sum_n\frac{\alpha^n}{\sqrt{n!}}\ket{n}.\\ \nonumber
\end{equation}
Projecting into the coordinate representation in two dimensions

\begin{align}\label{eq:cohstatfun}
\psi_\alpha(x,y) &=\braket{\mathbf{r}|\alpha_x,\alpha_y}\\ \nonumber
&= \sum_{n,m}\mathrm{A_{n,m}}\braket{\mathbf{r}|n,m}\\ \nonumber
&= \sum_{n,m}\mathrm{A_{n,m}}\mathrm{u_{n,m}}(x,y),
\end{align}
where $\braket{\mathbf{r}|n,m}=\mathrm{u_{n,m}}(x,y)$, 
$\alpha=\{\alpha_x,\alpha_y\}$, and the matrix elements of the coherent state are $\mathrm{A_{n,m}}=\alpha _x^\mathrm{n} \alpha _y^\mathrm{m}\exp\left(-\frac{1}{2}| \alpha _x|^2-\frac{1}{2}| \alpha_y|^2\right)/\sqrt{\mathrm{n! m!}}$. Now, the matrix elements of the state operator are 

\begin{equation}\label{eq:melemcohe}
    \mathrm{\rho^\alpha_{n,m,n^\prime,m^\prime}=A^*_{n,m}A_{n^\prime,m^\prime}}
\end{equation}

Then, in order to obtain the coherent vortex state, we multiply the state wave function ~\ref{eq:cohstatfun} by the core function $\mathrm{Z^{k}}(x,t,0)$. Therefore, after a straightforward calculation, using the recurrence properties of the Hermite polynomials, we can write the matrix elements of the coherent vortex state operator as

\begin{align}\label{eq:spancoh}
 \mathrm{A^{Z_k}_{n,m}}&=\frac{\mathrm{w}}{\sqrt{(x_\mathrm{k}-\sqrt{2}\mathrm{w} \Re(\alpha
   _x))^2+(y_\mathrm{k}-\sqrt{2}\mathrm{w}\Re(\alpha
   _y))^2+\mathrm{w}^2}} \\ \nonumber
&\times\left(\sqrt\frac{\mathrm{n}}{2} \mathrm{A_{n-1,m}} +i\sqrt\frac{\mathrm{m}}{2} \mathrm{A_{n,m-1}}\right. \\ \nonumber
&+\left.\sqrt\frac{\mathrm{n+1}}{2} \mathrm{A_{n+1,m}} +i\sqrt\frac{\mathrm{m+1}}{2} \mathrm{A_{n,m+1}}-(x_\mathrm{k}+iy_\mathrm{k})\mathrm{A_{n,m}}\right)\,,
\end{align}

where $(x_\mathrm{k},y_\mathrm{k})$ is the initial position $(t=0)$ of the vortex core, and $\Re$ stands for the real part of the coherent complex amplitude. Finally, the matrix elements of the coherent vortex state operator are

\begin{equation}\label{eq:melemcohvort}    \mathrm{\rho^{Z_k}_{n,m,n^\prime,m^\prime}=\left(A^{Z_k}_{n,m}\right)^*A^{Z_k}_{n^\prime,m^\prime}}
\end{equation}

A wavepacket with a phase defect results in a rotation of the quantum fluid around a vortex core. 

\end{document}